# Pure passive fiber enabled highly efficient Raman fiber amplifier with record kilowatt power

Yizhu Chen[1], Jinyong Leng[1], Hu Xiao[1], Tianfu Yao[1], and Pu Zhou[1]

[1] College of Advanced Interdisciplinary Studies, National University of Defense Technology, Changsha 410073, China

Corresponding author: Pu Zhou (e-mail: zhoupu203@163.com).

This work was supported in part by the National Natural Science Foundation of China (NSFC) (No. 11704409) and Huo Yingdong Education Foundation of China (No. 151062).

**ABSTRACT** Kilowatt-level high efficiency all-fiberized Raman fiber amplifier based on pure passive fiber is proposed for the first time in this paper. The laser system is established on master oscillator power amplification configuration while a piece of graded-index passive fiber is utilized as stokes shifting as well as gain medium, which is entirely irrelevant to rare-earth-doped gain mechanism. When the pump power is 1368.8 W, we obtained 1002.3 W continuous-wave laser power at 1060 nm with the corresponding optical-to-optical efficiency of 84%. The beam parameter $M^2$ improves from 9.17 of the pump laser to 5.11 of the signal laser through the amplification process, and the brightness enhancement is about 2.57 at maximum output power as a consequence of the beam clean-up process in the graded-index fiber. To the best of our knowledge, we have demonstrated the first kilowatt-level high efficiency Raman fiber amplifier based on pure passive fiber with brightness enhancement.

**INDEX TERMS** Fiber lasers, fiber nonlinear optics, high power amplifiers, Raman scattering.

## I. INTRODUCTION

High power fiber lasers (HPFLs) have been the subject of considerable attention over decades of years in commercial and civilian applications covering industrial manufacturing, material processing and medical treatment [1, 2]. Among these fiber lasers, stimulated Raman scattering (SRS) is one of the main physical limits which challenges power scaling, in that at high power level more signal light transfers into Raman light at longer wavelength and the output laser power declines [3]. From another point of view, Raman fiber lasers (RFLs) are wavelength-agile alternative to obtain high power output, which utilize SRS to obtain gain in optical fiber [4-9]. The generated wavelength is determined by the provided pump wavelength and stokes shift, therefore in RFLs there is wide achievable lasing waveband by controlling these parameters. When shorter pump wavelength and shifting range are chosen, the quantum defect is low which can mitigate the heat load [10]. In addition, generally in relatively long passive fibers with common doping concentration of germanium there is no photon darkening (PD) [11]. In the past decade RFLs with high power level have been demonstrated with potential in power scaling, and the typical structure is core-pumped RFL based on pure Raman gain which can realize several-hundred-watt level laser [6-8]. However in these laser systems the stokes shifting is achieved in fiber core, as a consequence it can hardly enhance the laser brightness through amplification. Then higher power output is acquired through fiber lasers employing integrated Yb-Raman gain [12-14], nevertheless the brightness decreases in the Raman shifting process and the issues of PD reappear which will influence the performance of laser system.

Recently the fast development of fiber components and the brightness enhancement of pump lasers make it possible for power scaling in passive fibers that employ pure Raman gain [15-22]. For instance, using graded-index (GRIN) fiber in core-pumped structure has a relatively simple construction and enhances the brightness in fiber core. In 2016, Y. Glick et al. report an LD-pumped Raman fiber laser in free space configuration which obtain 80 W Raman laser output, and brightness enhancement is achieved through the beam clean-up procedure of GRIN fiber in the cavity [18]. Then by means of modifying the system and adding up pump energy the lasing power increases to 154

W [19]. In 2017, A. Zlobina et al. propose an all-fiber Raman fiber laser that also utilizes GRIN fiber and acquire power of 50 W at 954 nm, while the FBGs inscribed in GRIN fiber are the key devices which are specially fabricated for efficient reflection and mode selection effect [20]. In 2018 Y. Glick et al. report an all-fiber Raman laser based on GRIN fiber and obtain power of 135 W at 1081 nm with brightness enhancement [21], and these achievements manifest that in pure RFLs the employment of GRIN fiber has potential in power scaling as well as brightness enhancement. It should be noted that the mentioned achievements are realized in oscillator configuration, which indicates that the handling and endurance ability of FBGs may be one of the restrictions for power scaling.

Apart from this, master oscillator power amplification (MOPA) configuration is another alternative for HPFLs that can acquire high laser power, and as a matter of fact most several-kilowatt HPFLs utilize this configuration at present time [22-27]. This architecture avoids the stability problems of high power oscillators, meanwhile the power scaling will benefit from the higher coupled pump power. Previously, we demonstrate an all-fiberized Raman fiber amplifier (RFA) employing GRIN fiber and obtain several-hundred-watt level high power output as well as brightness enhancement, which validates the feasibility of realizing high power laser in pure passive fiber enabled MOPA configuration [28]. In the present work, we report the first kW-level Raman fiber amplifier based on pure passive fiber. The laser system is established based on MOPA configuration, in which the signal laser is Ytterbium-doped fiber laser (YDFL) lasing at 1060 nm, correspondingly pump lasers at 1018 nm are utilized which match well due to typical Raman gain spectrum in silica fiber. In the amplifier GRIN fiber is employed as stokes shifting and gain medium. When the pump power reaches 1368.8 W, the total output power is 1268 W at maximum and correspondingly 1060 nm laser power is 1002.3 W with the optical-to-optical efficiency of 84%. The beam parameter $M^2$ improves from 9.17 of the pump laser to 5.11 of the signal laser through the amplification process, and the brightness enhancement is about 2.57 at maximum output power as a result of the beam clean-up process in GRIN fiber. To the best of our knowledge, we have demonstrated the first kilowatt-level high efficiency RFA based on pure passive fiber.

## II. EXPERIMENTAL SETUP

As is shown in Fig. 1, the experimental design of the amplifier is based on MOPA mechanism. One single mode YDFL operating with central wavelength of 1060 nm serves as the seed laser which emits 75 W signal light, while six home-made high power YDFLs lasing at 1018 nm offer pump energy [29]. The Ytterbium-doped fiber (YDF) of these lasers has the same core/inner-cladding diameter of 15 μm/130 μm, and the laser cavity parameters have been optimized for good performance. The available pump energy is 1368 W in the aggregate and all the YDFLs are coupled into a home-manufactured 7×1 combiner [30]. The radial size of input fiber is the same with the pigtail of the YDFLs, while the core and cladding diameter of output fiber are 50 μm and 360 μm, respectively. A piece of double-clad GRIN fiber is spliced to the output fiber of the combiner that serves as Raman convertor and gain medium, which has length of 50 m with core/cladding diameter of 62.5 μm/125 μm. The refractive index profile of the GRIN fiber is measured and the core NA is around 0.275. At the end of the GRIN fiber an end cap is employed for high power lasing. To acquire better cooling effect, the GRIN fiber winds around a metal barrel and the whole system is placed on water-cooling plate which maintains at 18℃.

The performance of the fiber amplifier is monitored through the beam splitting system and detecting system as is shown in Fig. 1. The beam splitting system consists of a set of optical lens, including one collimator mirror (CO), one dichroic mirror (DM) and one high reflectivity mirror (HR). The exit laser output from the end cap consists of signal light at 1060 nm and residual pump light at 1018 nm. After the end cap the laser is first collimated by CO, then by utilizing DM the residual pump light is reflected while the signal laser is transmitted. The HR has reflectivity of over 99%, which can split up a fraction of signal laser for beam quality measurement. In the detecting system the laser power is recorded by power meter, while the wave spectrum is measured by optical spectrum analyzer (YOKOGAWA, AQ6370D) through monitoring the scattering light of output laser. The beam quality parameter and beam shape at focal spot are measured by BQ analyzer (Laser Quality Monitor from Prime Company).

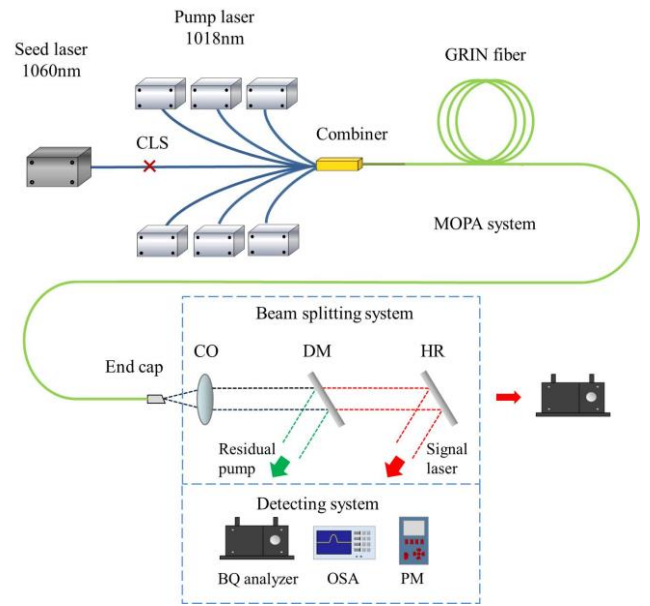

**FIGURE 1.** Schematic of Raman laser amplifier based on GRIN fiber, the beam splitting system and detecting system.

## III. RESULTS AND DISCUSSIONS

Fig. 2 demonstrates the power scaling characteristics of the Raman fiber amplifier. The amplification process begins when pump light at 1018 nm is injected, and the signal power at 1060 nm starts to increase. In that at the beginning the amplification is not sufficient, the growth of signal power is relatively slower than that of residual pump laser at 1018 nm. Then with the increment of pump power, the growth of signal laser power accelerates and the scaling process is almost linear. When the pump power is 1368.8 W, the total output laser reaches 1268 W at maximum, of which the 1060 nm laser power is 1002.3 W and the optical-to-optical efficiency is 84%. Further increase of signal laser is mainly limited to available pump energy. The residual pump power increases at first, then with the rapid amplification the pump power gradually decreases to 265.7 W.

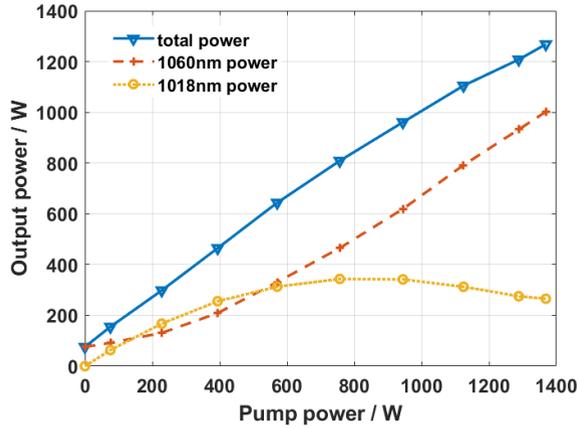

FIGURE 2. The output power characteristics (including total output power, 1060 nm laser power, and residual 1018 nm laser power measured from end cap).

The output spectrum of Raman amplifier on logarithmic scale is monitored as is shown in Fig. 3. With the increment of pump power, the signal laser keeps increasing. On account of the gain competition of amplified spontaneous emission (ASE), there is laser generating at 1025 nm ~ 1045 nm in pump lasers [31], which is also indicated in the spectrum in Fig. 3. Despite of this, our amplifier functions with stable lasing and no self-oscillation appears. High order Raman light at 1113nm arises in the spectrum when the signal laser power reaches 331 W, and at this time the signal wave spectrum strongly broadens and lateral wings are observed on both sides of the signal spectrum which results from four-wave-mixing (FWM) [32]. Then the power of signal and high order Raman wave keep increasing, and the full width at half maximum (FWHM) of signal spectrum increases from 0.66 nm (seed laser) to 1.14 nm (maximum output power). Under maximum output signal power the high order light is about 30 dB lower than signal light of which the power is negligible compared to signal laser.

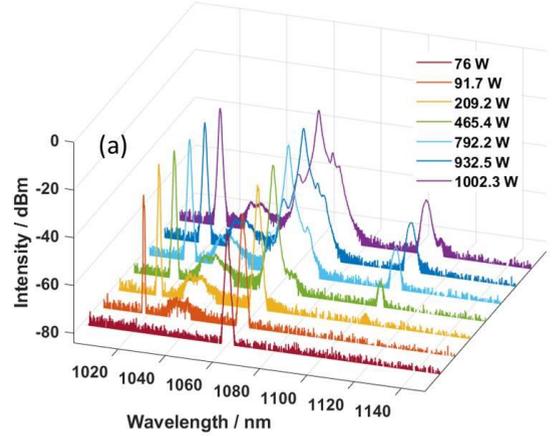

FIGURE 3. Spectrum under different launching pump power levels in logarithmic scale(legend is the corresponding output signal laser power).

Through our detecting system the beam quality parameter is also monitored, as is illustrated in Fig. 4. The $M^2$ of pump laser is first measured after the end cap, while the BQ remains stable and $M^2$ slightly increases from 9.17 (75.7 W) to 9.8 (1288 W). In the amplification the $M^2$ of signal and residual pump laser are also recorded. During the power scaling procedure, the $M^2$ of signal laser decreases as a result of beam clean-up scheme in GRIN passive fiber. On the contrary, the $M^2$ of residual pump laser keeps increasing, which may indicate that in pump laser the proportion of light with good beam quality converts to the signal laser. This outcome can be concluded from the transverse distribution of pump and Raman light as well as the overlap integral of different modes in the GRIN fiber [33, 34].

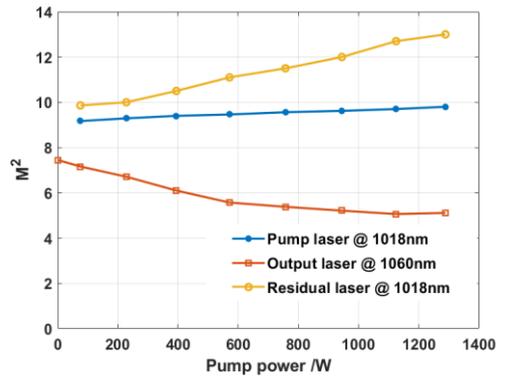

FIGURE 4. $M^2$ of pump laser, residual pump laser and output signal laser vary with increasing pump power.

Fig. 5 illustrates the beam spot at focal point recorded by the BQ analyzer. The beam quality measurement results of pump light and seed laser after the combiner and the end cap are shown in Fig. 5 (a) - (d), respectively. The $M^2$ of pump light and seed laser both rise up after transmitting in the GRIN Raman fiber and beam shape of the laser degrade. Through the amplifying process the $M^2$ of signal laser decreases from 7.45 to 5.11 gradually, as is demonstrated in Fig. 5 (e) - (h). A portion of laser leaks off from the cladding of the GRIN fiber as a result of manufacturing defect which may influence the measurement of beam quality, nevertheless the beam quality shows an ascending tendency as a consequence of beam clean-up process.

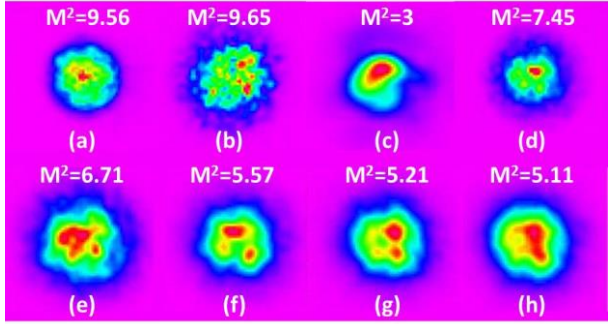

**FIGURE 5.** Beam shape at focal spot measured by BQ analyzer of pump light from combiner (a) and end cap (b), seed laser from combiner (c) and end cap (d), and amplified signal laser from end cap with power 209.2 W (e), 465.4 W (f), 792.9 W (g), and 1002.3 W(h).

On the basis of beam quality measurement, we calculate the brightness enhancement of the fiber amplifier which is illustrated in Fig. 6. At the beginning of the amplification the brightness enhancement is 1.83, and with the increase of pump power the brightness enhancement decreases to some extent then keeps increasing again to 2.57. The formation of the notch of the curve is related to the amplification efficiency. With more injected pump light the brightness of pump laser continues increasing, while at the beginning of the amplification the signal laser power rises up with lower growth rate than pump light as is shown in Fig. 2. Consequently at this time the increment of brightness of signal laser at 1060 nm is slower than that of pump laser at 1018 nm, which means the brightness enhancement (quotient between brightness of signal and pump laser) decreases. When the amplification is more sufficient, the Raman conversion efficiency increases and remains stable, which manifest that more pump light converts to signal laser. Therefore the brightness of signal light rises more rapidly than pump laser, and the brightness enhancement continues increasing. The contrast of 3D isometry of the beam shape at focal spot indicates the beam clean-up effect.

Before this experiment GRIN fiber of different fiber length are utilized to test the performance of the Raman amplifier including 400 m, 300 m, 250 m and 100 m. The power scaling of these fiber amplifiers are limited to the high order Raman threshold, and the correspondingly maximum signal output power at 1060 nm is 75.9 W, 121.7 W, 134.5 W and 528.8W. Through optimizing the length of GRIN fiber, the high order Raman threshold is improved and the obtainable maximal signal laser power increases. It can be predicted that higher lasing power can be achieved if further majorization of other parameters in GRIN fiber is realized in this configuration. Meanwhile, in all these laser systems the beam quality keeps increasing and shows potential in brightness enhancement, which can acquire optimization in future experiment.

## IV. CONCLUSION

In general, we report the first kW-level pure Raman gain enabled fiber amplifier. The laser system is established based on MOPA configuration, in which the signal laser lasing at 1060 nm and pump lasers at 1018 nm are utilized that match well related to typical Raman gain spectrum in silica fiber. A piece of GRIN Raman fiber is utilized as stokes shifting and gain medium in the amplification. When the pump power is 1368.8 W, the total output power reaches 1268 W at maximum, of which the 1060 nm laser power is 1002.3 W and the corresponding optical conversion efficiency is 84%. The beam parameter $M^2$ improves from 9.17 of the pump laser to 5.11 of the signal laser through the amplification process, and the brightness enhancement is about 2.57 at maximum output power as a consequence of the beam clean-up process in GRIN passive fiber. To the best of our knowledge, this is the first demonstrated kilowatt level high efficiency Raman fiber amplifier based on pure passive fiber.

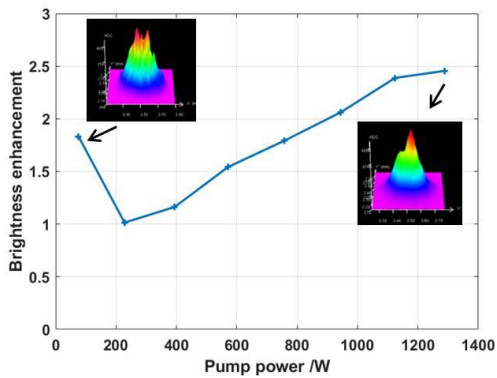

**FIGURE 6.** The brightness enhancement varies with the increasing pump power (the sub-graph is the 3D isometry of the beam shape at focal spot).